\documentclass[10pt,twocolumn,letterpaper]{article}

\usepackage
[
        letterpaper,
        left=.75in,
        right=.75in,
        top=.75in,
        bottom=.75in,
]{geometry}

\usepackage{amsmath}
\usepackage{xcolor}
\usepackage{amssymb}
\usepackage{wasysym}
\usepackage{titlesec}
\usepackage[font=small,labelfont=bf]{caption}
\usepackage{subcaption}
\usepackage{times}
\usepackage{url}
\usepackage{graphicx}
\usepackage{wrapfig}
\usepackage{paralist}
\usepackage{enumerate} 
\usepackage{enumitem}
\setlist{nolistsep}
\usepackage{sidecap}
\usepackage{adjustbox}
\usepackage{tikz}
\usepackage{multirow}
\usepackage{pgfplots}
\selectcolormodel{cmyk}
\usepackage{pgfplotstable}
\usepackage{chronology}

\usepackage[biblabel]{cite}

\usepackage{tikz}
\usepackage{pgfplots}
\usepgfplotslibrary{fillbetween}
\usepgfplotslibrary{groupplots}
\pgfplotsset{compat=1.9}
\usetikzlibrary{pgfplots.dateplot}
\usepackage{pgfplotstable}

\titleformat{\section}{\large\bfseries}{\thesection}{1em}{}
\titleformat{\subsection}{\itshape}{\thesubsection}{1ex}{}{}
\titleformat{\subsubsection}{\itshape}{\thesubsubsection}{1ex}{}{}
\titlespacing{\section}{0pt}{*1.2}{*.5}
\titlespacing{\subsection}{0pt}{*1.5}{*.3}
\titlespacing{\subsubsection}{0pt}{*1}{*0}
\titlespacing{\paragraph}{0pt}{3pt}{*1}

\newcommand{\ie}{\textit{i.e.}}

\newcommand{\nop}[1]{}

\usepackage{mathtools}

\newcommand{\hl}{}

\newcommand{\hlnew}{}

\usepackage{authblk}

\begin{document}

\title{Behavior Change in Response to Subreddit Bans and External Events}

\author[1]{Pamela Bilo Thomas}
\author[1]{Daniel Riehm}
\author[2]{Maria Glenski}
\author[1]{Tim Weninger}

\affil[1]{Department of Computer Science \& Engineering,  University of Notre Dame}
\affil[ ]{{\small\texttt{\{pthomas4, driehm, tweninge\}@nd.edu}}}
\affil[2]{Data Sciences and Analytics Group, Pacific Northwest National Laboratory}
\affil[ ]{\small\texttt{maria.glenski@pnnl.edu}}

\renewcommand\Authands{, and }

\date{}

\twocolumn[
  \begin{@twocolumnfalse}
    \maketitle
    \begin{abstract}
As more people flock to social media to connect with others and form virtual communities, it is important to research how members of these groups interact to understand human behavior on the Web. In response to an increase in hate speech, harassment and other antisocial behaviors, many social media companies have implemented different content and user moderation policies. On Reddit, for example, communities, \ie, subreddits, are occasionally banned for violating these policies. We study the effect of these regulatory actions as well as when a community experiences a significant external event like a political election or a market crash. Overall, we find that most subreddit bans prompt a small, but statistically significant, number of active users to leave the platform; the effect of external events varies with the type of event. We conclude with a discussion on the effectiveness of the bans and wider implications for the online content moderation.
\vspace{.7cm}
    \end{abstract}
  \end{@twocolumnfalse}
]

\section{Introduction}\label{sec:introduction}
Users of online social platforms tend to self-organize into communities. These user-defined communities are a large component of online discourse where users express their ideas and share content about current events. Usually, online communities are organized around a topic or common theme and are governed by rules of behavior and etiquette~\cite{fiesler2018reddit}. 

\hl{We consider the case of Reddit,} where online social communities called subreddits commonly sprout up in support of sports teams, movie franchises, video games, political causes, and so on. These communities exist to support and discuss various aspects of the topic with respect to current events. Frequently, external shocks occur which dramatically challenge the identity of the community. Consider, for example, the subreddit in support of the Rams, a professional American NFL football team which moved its home from St. Louis to Los Angeles in the Spring of 2016. When that event occurred, the online community experienced significant upheaval: a new subreddit, r/LosAngelesRams, was created and r/StLouisRams experienced a marked decline in activity. This example raises several socially relevant questions: how many users followed the Rams to the new community? Did existing users disengage from the Rams? and if so, did they join a different fandom? How many Rams-fans left Reddit entirely?

\hl{In other cases, there are some subreddits that have been deemed toxic to the larger Reddit community, and have been banned for violating Reddit's community standards. These regulatory actions are known to have a variety of impacts on the respective communities, including changes in posting activity, movement to other communities, or leaving the site altogether.}

External and regulatory shocks like these occur frequently in online social communities. Although the story of the NFL's Rams is important to millions of people, the goal of the present work is much broader. We view these events as opportunities to study human behavior and community change as natural experiments. Under perfect conditions, natural experiments like these can uncover causal effects of community and user behavior in response to events. Unfortunately, perfect conditions are rare and confounds are likely present in our data -- a fact that we will revisit in the methodology section of this paper.

\hl{In the present work, we explore the effect of these external and regulatory shocks on both the individual and community level. To do this, we identified a variety of communities that have been through shocks and track their users to see what effect the shock had on their posting behavior.  Specifically we are interested in answering the following research questions:}


\noindent\textbf{RQ1: How do online communities react to different kinds of shocks?} The move of an NFL team (external event) is certainly different from an outright ban of a subreddit (regulatory action). By comparing subreddit bans against external events we can better understand the relative effect of these bans, and we can determine if they are more or less disruptive than the effect of external events. 

\noindent\hl{\textbf{RQ2: How do individual users react to these shocks?} Users of online social systems may react differently to various external events. To answer this question, we analyze which users change their behavior as a result of the ban and attempt to create a model that is capable of predicting whether or not a user will leave the site.}

\noindent\hl{\textbf{RQ3: How many users change their behavior as a result of these shocks?} We use Bayesian Changepoint Analysis to investigate how many users underwent a behavior shift as a result of a shock. These changes could be represented as an increase or decrease in posting behavior, or a change in which subreddits these users are posting in.}


\hl{Overall, we find that most Reddit bans prompt a small, but statistically significant (one-tailed $p < 0.05$) number of affected users to leave Reddit (approx. 4\% to 10\%) while the effect of external events varies with the type of event.}  

\section{Related Work}

\subsection{Digital Traces}

\hl{
Digital traces produced by user activity in online social spaces allow for detailed observational studies of computer-mediated human behavior.} \hlnew{The visualisation of social media analytics is especially important~\cite{chen2017social}.} \hlnew{Location-based social media, for example, has been a helpful way of understanding how events affect various communities~\cite{dunkel2019conceptual}}. \hl{Fine-grained studies in community construction~\cite{ozer2016community}, inter-group conflict~\cite{datta2019extracting}, intra-group coordination~\cite{kraut2012building,weninger2014exploration}, and attention~\cite{tan2015all,zhang2017community} have also been made possible by the recent availability of these digital traces of online behavior. Analysis of user behavior has been done on other social media sites such as Pinterest~\cite{gilbert2013need}, Quora~\cite{wang2013wisdom},} \hlnew{WeChat~\cite{li2018weseer}, and Twitter~\cite{lu2014integrating}.  In one case, the box office success of a film could be predicted using chatter about the movie on social media~\cite{lu2014integrating}, and information cascades were able to be accurately forecast using data from WeChat~\cite{li2018weseer}}. \hlnew{ On Reddit in particular, users are have defined to be either content-producers or content-consumers~\cite{thukral2018analyzing}}. Reddit has been used in other social media tasks, such as modelling online comment threads~\cite{krohn2019modelling} and predicting user interactions~\cite{glenski2017predicting}. \hlnew{Oftentimes, reactions to events can cause subsequent events in various communities~\cite{dunkel2019conceptual}}.

\subsection{Reddit Communities}

\hlnew{Continuing our focus on Reddit, we note that many subreddits are driven by user retention and growth -- by creating an environment where posters and commenters feel welcome and included. To that end, Joyce and Kraut defined three ways by which people continue to post in a community: (1) reinforcement, where a user has positive reactions with others, (2) reciprocity, where a user does something for someone else, and the other returns the favor, and (3) through personal bonds with members~\cite{joyce2006predicting}. Others have found that stability, cohesiveness, and sociability are important to the future of online communities~\cite{mcewan2016communication,brandtzaeg2008user}. As communities grow, participation in group discussion tends to become more concentrated~\cite{panek2018effects}. 
Strong connections to parent subreddits, defined as subreddits that spawn new subreddits, lead to stronger growth in communities~\cite{tan2018tracing}. When two communities share users, they tend to be excited about the same things, and interest intrinsically springs up in both groups~\cite{backstrom2006group}.}

\subsection{Online Toxicity}

\hl{The freedom of these user-defined communities which provides a space for digital fandoms to flourish also gives way to toxic communities that promote hate speech~\cite{chandrasekharan2017you,saleem2017web,brandtzaeg2008user} and misinformation~\cite{starbird}. Communities address toxic behavior through regulations that constrain activity in different ways~\cite{lessig2009code}. For example, banning subreddits that violate Reddit's anti-harassment policy has shown to effectively reduce the occurrence of hate speech on the platform~\cite{chandrasekharan2017you}; however, users of banned communities may simply move to less-regulated parts of the Web~\cite{newell2016user}}.

\hlnew{Machine learning models have been created to predict if a subreddit will become hateful or toxic~\cite{habib2019act}. Likewise, predictive models have been created that can indicate whether or not a conversation is about to turn uncivil~\cite{zhang2018conversations} or continue to grow in popularity~\cite{cunha2019all}. It is difficult to simply try to find hateful communities by using keyword search, because, counter-intuitively, words that appear in hateful communities are also used in supportive communities~\cite{saleem2017web}. Instead, researchers have developed a `bag of communities' approach which compares hate-filled subreddits to well-moderated subreddits in order to classify other communities as hateful or not~\cite{chandrasekharan2017bag}. }

\subsection{Changepoint Analysis}

As an additional part of our study, we look to use changepoint analysis to uncover more behavior change.  \hl{Besides social media post behavior, changepoint detection has been used in a wide variety of other areas, such as detecting changes in individuals who are at-risk for suicide attempts by analyzing their Twitter posts~\cite{vioules2018detection}, anomalies in race car driving~\cite{widanage2019anomaly}, and sociopolitical events based upon historical documents~\cite{chaney2016detecting}}.

\section{Data and Methods}

\subsection{Research Framework}

We focus our analysis in the present work on Reddit. Users on Reddit self-organize their discussions into topic-based communities called subreddits. By default, posts and comments are sorted by popularity; posts and comments with many (up)votes are viewed by more users, which then have an opportunity to further rate the post or comment~\cite{glenski2017consumers}.  

\hl{Reddit is an almost ideal use case for our analysis. Information posted to Reddit is publicly available. This data includes the username, the timestamp, and content of the comment, as well as the community in which the comment was posted. Unlike Twitter and Facebook, comments on Reddit are threaded, which provides a simple and intuitive structure from which further information can be gleaned.  Additionally, the subdivision and self-categorization of subreddit-communities allows for focused analysis.} Community membership and activity evolves over time, rising and falling over time as user interests and perceptions change. Identity-based communities (as opposed to humor or image sharing communities) tend to grow quickly~\cite{barbosa2016averaging} and have high rates of retention~\cite{farzan2011increasing}. But shocks to this stability sometimes occur. In the present work, we study two types of shocks that occur on Reddit: 1) external events and 2) regulatory actions.

%

\hlnew{Although the present work focuses on Reddit, the paper may serve as a general framework for the analysis of the effects of external events and regulatory actions on user participation in online social spaces. The events assessed in the present work provide opportunities to perform quasi-experiments to better understand the causal impacts of these externalities on online social systems. Although there are only a handful of popular subreddits that have been banned, there are indeed many subreddits that have experienced external shocks. We only sample a few in the present work. Specifically, our research methodology includes Bayesian changepoint analysis, a difference in differences test which measures changes in user activity, and an analysis of user retention. This framework provides an intuitive look at how social media users continue or change their social media behavior in the wake of a subreddit ban or significant external event.}



\subsection{Quasi Experiments}
Much of the work in online social analysis relies on observational methodologies to elicit descriptive statistics. These observational studies are valuable, however many of our questions can only be addressed with causal methods. Unfortunately, randomized controlled experiments of how online users respond to various social stimuli are rare because of the difficulty involved in implementing treatments in massive online social systems.

In an ideal randomized controlled experiment designed to elicit casual effects, a subreddit would have been randomly chosen and either banned it or encountered some topically relevant external shock. Unfortunately, neither of those scenarios is likely to occur, so we use a technique to approximate this ideal scenario. These include 1) the creation of control groups using two kinds of user matching and 2) using a difference-in-differences methods to compute the effect size relative to the control group.


\section{Reddit Reactions}

\subsection{External Events}

We define an external event as an event \nop{Im not sure here... this language is tough... occurance? lets come back to it} that occurs outside of the digital community that suddenly and dramatically changes the identity of the group. Our goal is to understand how these online communities react to these external events and, later, compare the effect external events with the effect of regulatory actions. 

For example, we consider behavioral differences between the r/StLouisRams and r/Chargers subreddits in reaction to each team's relocation. The Rams and the Chargers are two professional American football teams who compete in the National Football League (NFL) who each moved their home to Los Angeles within relatively quick succession, with the Rams moving in 2016 and the Chargers moving in 2017. In this case, the Chargers fans kept their subreddit and Rams fans created a new subreddit: r/LosAngelesRams. Comparing the user retention between these communities is interesting as we compare how these communities deal with the external shock of losing their football team.

Similarly, we analyze how r/StarWars reacted to the opening of the two new films in the ``sequel'' trilogy including ``The Force Awakens'' (2015) and ``The Last Jedi'' (2017). Did StarWars community-members change their behavior in response to these external events? And if so, how?

Finally, we analyze how members of r/SandersForPresident and r/The\textunderscore Donald reacted to the 2016 US Democratic and Republican primary elections, in which Sanders did not win and Trump did win, as well as the US general election, which Trump won and Sanders did not participate. Using these examples, we show how politically oriented community members change their behavior after their candidate wins or loses an election.

\subsection{Regulatory Actions}

We define a regulatory action as an action taken by the online social platform to limit or eliminate the behavior of a user or community. 
Each subreddit is regulated by moderators, who develop community guidelines~\cite{fiesler2018reddit,chandrasekharan2018internet} and police their communities for nonconforming behavior~\cite{birman2018moderation}. Uneven community standards coupled with Reddit's hands-off regulation had permitted hate speech to flourish in previous years. Additionally, this lack of regulation has allowed for toxic and offensive content to spread through many subreddits~\cite{massanari2017gamergate},  which has been shown to depress subreddit growth~\cite{mohan2017impact}. 

In May 2015, following years of controversy about their hands-off approach to regulating speech on their platform, Reddit began to implement an anti-harassment policy. This led to bans of subreddits such as r/fatpeoplehate and r/CoonTown, which had thousands of subscribers and active members at the time of the ban. The r/fatpeoplehate subreddit was banned on June 10, 2015, due to repeated violations of Reddit's hate speech and harassment regulations. After the ban, many users who formerly posted to r/fatpeoplehate created new forums to continue their activity, but these were quickly banned as well; other users left Reddit altogether. Those who stayed used less hate speech in their subsequent activity~\cite{chandrasekharan2017you}. However, it is likely that many affected users simply moved elsewhere~\cite{chandrasekharan2017you,newell2016user,saleem2018aftermath}.

Since this event, Reddit has gone on to ban other troublesome subreddits from its platform. Many of these subreddits are highlighted in Table~\ref{tab:sub_data}.
%


\begin{table}
\caption{Subreddits that experienced external events or regulatory actions. Active users are those who have posted or commented at least 10 times in the year prior to the event.}
\centering
\scriptsize{
\begin{tabular}{l||r|l|l|l}
\multicolumn{5}{c}{\footnotesize{Regulatory Actions}} \\
\textbf{Subreddit} & \textbf{Date} & \textbf{\# Active} & \textbf{Subscribed} & \textbf{Event} \\ \hline
r/Physical\textunderscore Removal&15-Aug-17&576&9541&Ban\\ \hline
r/GunsForSale&21-Mar-18&1040&23082&Ban\\ \hline
r/SanctionedSuicide&14-Mar-18&1930&15672&Ban\\ \hline
r/DarkNetMarkets&21-Mar-18&5295&175760&Ban\\ \hline
r/Incels&7-Nov-17&6685&42236&Ban\\ \hline
r/fatpeoplehate&10-Jun-15&16548&151404&Ban\\ \hline
r/greatawakening&12-Sept-18&7605&70660&Ban\\ \hline
r/MillionDollarExtreme&10-Sept-18&5924&35400&Ban\\ \hline
\multicolumn{5}{c}{} \\
\multicolumn{5}{c}{\footnotesize{External Events}} \\
\textbf{Subreddit} & \textbf{Date} & \textbf{\# Active} & \textbf{Subscribed}  & \textbf{Event} \\ \hline
\multirow{2}{*}{r/The\textunderscore Donald}&18-Jul-16&35615&176294&RNC Conv.\\
&9-Nov-16&62251&280092&US Elect.\\ \hline
r/SandersForPresident&26-Jul-16&40719&225728&DNC Conv.\\ \hline
r/StLouisRams&1-Jan-16&723&4529&LA Move\\ \hline
r/Chargers&1-Jan-17&1513&11497&LA Move\\ \hline
\multirow{2}*{r/StarWars}&18-Dec-15&12435&355089&Ep7\\
&15-Dec-17&20017&731707&Ep8\\ \hline
r/DarkNetMarkets&17-Dec-17&5963&167428&BTC crash\\
\end{tabular}
}
\label{tab:sub_data}
\end{table}

\subsection{Treatment Group}
For each subreddit we captured the post and submission history for the year prior and the year after the event. We call these {\em treatment-subreddits}. Unless otherwise indicated, we define an {\em active user} of a subreddit at the time of the shock as any user who posted at least 10 times in the year prior to the shock event. This default setting approximates the ``monthly active user'' widely used in Web and social media measurements; however, we find that high-, medium-, and low-activity users do result in different behavior. An active user of a treatment-subreddit is therefore called a {\em treatment-user}. 
%

\subsection{Control Groups}
We compare the behavior of the treatment group against two control groups (A and B). \hl{Our goal in identifying these control groups is to find matches to treatment-users so that causal conclusions can be made}.

Control group A is constructed using the methodology developed by Chandrasenkharan et al~\cite{chandrasekharan2017you} which identifies control-users as those who post in similar subreddits with a similar frequency. Following this methodology, we first define a {\em control-subreddit} to be all subreddits that a treatment-user posted to in the year prior to the event except the treatment-subreddit. 
We then refine this set of control-subreddits to include only those that have at least 10 posts by at least 10 treatment-users, \ie  we only include subreddits where treatment-users were active contributors.
We further refine the set of control-subreddits to include only the top 200 with the largest percentage of treatment-users. 

A potential {\em control-user} is therefore any user who was active (posted at least 10 times in the year prior to the event) in any control-subreddit but who has never posted in the treatment-subreddit.
Finally, for each treatment-user, we find a ``matching'' control-user from the set of potential control-users using the Mahalanobis Distance Matching algorithm according to the logarithm of the user's account age, total karma, and number of posts made in the last year. 


Control group B is constructed using the same methodology as control group A, except instead of curating a list of users who post in subreddits similar to the banned subreddit, we only consider those who post at least 10 times in r/AskReddit in the year prior to the event. This restriction identifies control-users who are active users on Reddit and match those that were a part of the banned community but who are less likely to be posting in hateful communities in the case where the treatment group consists of banned subreddits. Again, for each treatment-user, we compute the Mahalanobis distance for each AskReddit user to find a matching set of control-users.

\subsection{Reaction Measurements}
To measure how users react to community shocks like subreddit bans or other external events we plot various measures of user activity starting from year prior to the event until a year after the event. These plots demonstrate the change, if any, relative to the controls and therefore indicate the level of change in user behavior (approximately) caused by the event. 

Users may react in various ways to an event. First, the overall activity with the subreddit may change. We can measure activity in the raw number of posts and comments or by the number of unique users making a post or a comment. 

When a subreddit is banned, users of that subreddit are (typically) not banned themselves. They may remain on the site but focus their attention on other subreddits instead. Often, alternative subreddits are created in the immediate aftermath of a ban, but these new replacement subreddits are typically quickly shut down so we do not study that behavior in the present work. 


We measure the effects of the regulatory and external shocks through a temporal analysis of user posting and commenting behavior. We begin by separating user posting behavior into pre-shock and post-shock periods. We then bin user activity posts into time-windows of 7-day intervals. Our analysis is performed on the mean-averages of user-activity across the bins.

\subsection{Data}

We collected all available posts and comments on Reddit beginning from its inception in December of 2005 through September of 2019 from the Pushshift monthly archives of Reddit data\footnote{Archived Reddit data collected from \url{https://files.pushshift.io/reddit/}}. In total the dataset includes 733,160,370 posts and 6,170,787,222 comments making this one of the largest and most complete online social community datasets available. If a user, comment, post, or subreddit was deleted before the data ingestion began, then that content is not present in our dataset, which is important to remember as we consider the effect on banned communities. Similarly, posts and comments are collected soon after they are created so updates to posts or comments may not be captured.


\section{User Reactions}

Here we explore the user-level effects of events and bans (\textbf{RQ1}). We ask: What was the change in number of users? What was the change in overall activity level, \ie, number of post and comments, and type, \ie, where they post? And how does this activity compare between external and regulatory shocks?

\subsection{Attrition} 

We first consider the effect that external shocks have on user attrition. That is, in the event of a subreddit ban or external event, how many users leave Reddit altogether? Previous work has shown that subreddit bans have an especially large effect on user attrition~\cite{newell2016user,chandrasekharan2017you}, but these studies only look a few weeks into the future.

\begin{figure}[t!]
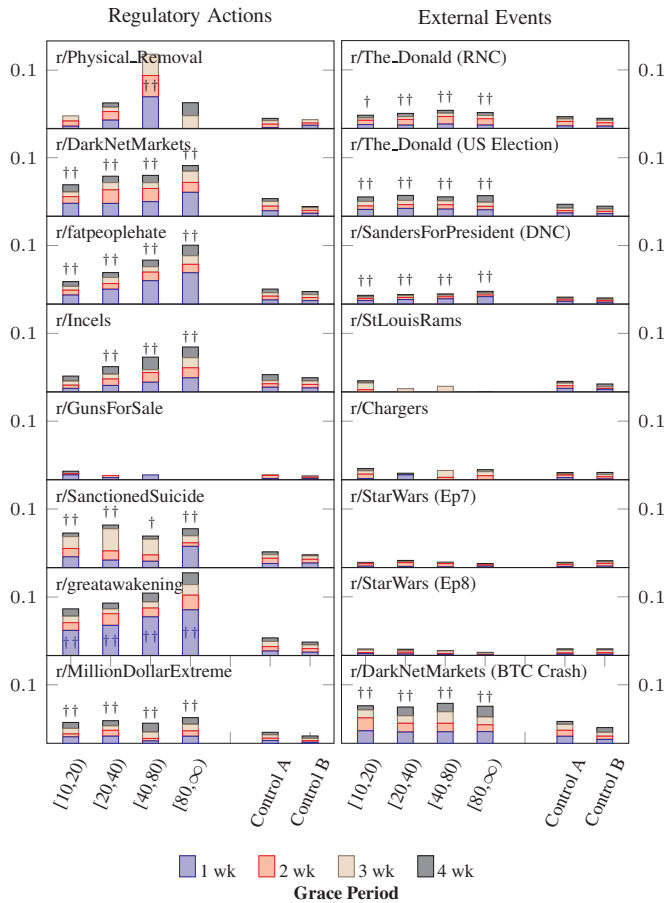

\centering
\include{figs/likelihood_of_leaving}
\caption{Rates of user attrition by number of posts in the subreddit or control group. Stacked bars indicate users that are inactive on Reddit after 1, 2, 3, and 4 week grace period. Left and right $\dag$s over each bar indicate treatment subreddits with attrition rates that are significantly higher (one-tailed $p<0.05$) than the attrition rates of Control A and Control B respectively.}
\label{fig:leaving}
\end{figure}

Here we define an inactive user as one who ceases to post or comment. Users may continue to visit the site, however, or upvote and downvote posts and comments but that data is not publicly available. In certain circumstances a user may wish to remain for some time and then leave the site, so we repeat our analysis after factoring in a grace-period of 1, 2, 3, and 4 weeks of residual activity before measuring for inactivity. For example, a user of a banned community may complain about the ban in another subreddit for a few days before eventually leaving the site and ceasing all activity. In this case, the short grace period permits this user to be counted as inactive. 

Figure~\ref{fig:leaving} illustrates the percentage of users who become inactive in the weeks following the ban or external event. The bar plots further break down the attrition analysis according to the number of posts in the year before the shock. 
We compare the attrition rates of treatment users against the natural attrition of Reddit users in each of the control groups. The attrition rates for many, but not all, shocks were found to be statistically significant using two-sample tests for equality of proportions. Attrition rates that were found to be statistically different from the controls ($p<0.05$) are indicated with a left and right $\dag$ symbol above the bar plots in Fig.~\ref{fig:leaving} representing statistical significance compared to Control A and Control B respectively. $p$-values are calculated using a two-sided test, but all significant results were from the right side of the distribution; that is, all significant results indicated more attrition compared to the control(s).

Although our previous analysis often found statistically significant user attrition, the overall effect was almost always less than 10\%. In other words, at least 90\% of users active in subreddits experiencing external shocks or bans continued to post and comment on Reddit. The effect-size was larger for bans on average, but not as much as we expected considering related findings~\cite{chandrasekharan2017you}.

\noindent\textbf{External shocks.} \hl{We see that in the case of the NFL teams and the Star Wars subreddits, there was not a significant number of users who left Reddit. However, we do find movement away from Reddit in many of the political subreddits.  We can hypothesize that the victory of a candidate caused their supporters to leave Reddit after the election season, because there was no longer a campaign to discuss. We see that in Table~\ref{tab:did}, there was a statistically significant decrease in users who frequented the StLouisRams subreddit, as compared to the Chargers subreddit.  This suggests that even though few individuals left Reddit, there was still a decrease in posting activity, perhaps because the change in team subreddits left many people behind}.

\noindent\textbf{Regulatory shocks.} \hl{Most regulatory shocks, or bans, caused a statistically significant number of individuals to leave the platform for most subreddits. One notable exception is the GunsForSale subreddit, which did not show significant attrition.  In the case of r/GunsForSale, active gun communities still continued to exist on Reddit, in r/Guns and r/GunAccessoriesForSale (which was created on the day of the r/GunsForSale ban).  Perhaps this is why more active users in r/GunsForSale continued to stay on the platform after the ban.}

\hl{In general we show that the overall effect of a ban was much larger than the effect of external events. We can hypothesize that users of subreddits that experienced an external shock still found Reddit to be a place to go to for discussion, especially since the shocked subreddit continued to exist.}

\subsection{Activity Change}

\begin{figure}[t]
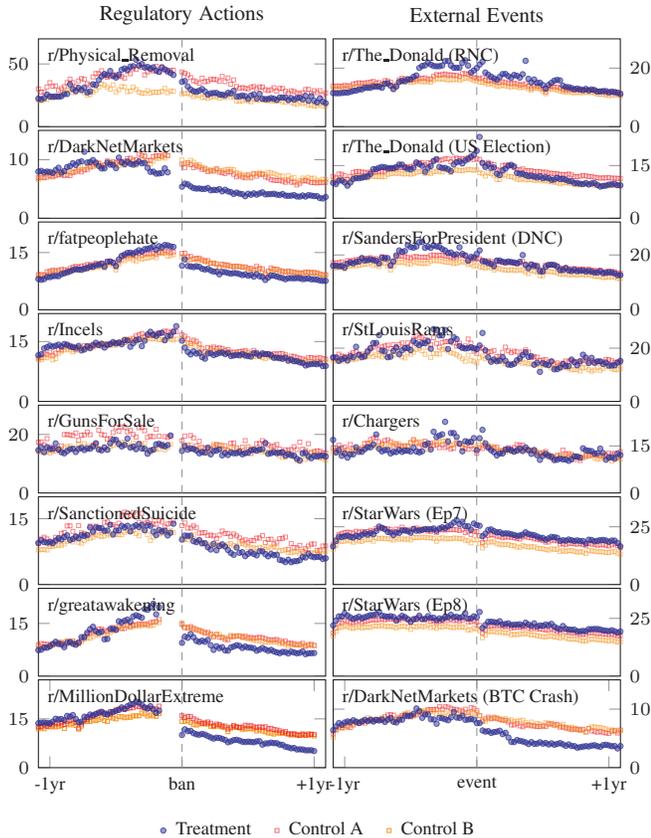

\centering
\include{figs/activity_bannedsubreddits}
\caption{Average user activity per week for the treatment group (\textcolor{blue}{\Circle}) compared to control group A (\textcolor{red}{\Square}) and control group B (\textcolor{orange}{\Square}) across all of Reddit. Timescales are centered on the date of the ban or external event and contain activity spanning one year before and after the event. Note that all y-axes start at 0, but their max is not consistent across all subplots. }
\label{fig:activity}
\end{figure}

Our next analyses focus on the activity of users that remained active on Reddit. First, we explore the change in the \textit{amount} of user activity using a difference in differences test. Second, we use multivariate Bayesian changepoint analysis~\cite{adams2007bayesian} to look for changes in the \textit{trends} of user activity, \hl{where a change is defined as an increase or decrease in posting or commenting behavior or a change in the communities to which the activity is targeted}.

The plots in Fig.~\ref{fig:activity} illustrate the average weekly activity for active users compared to the two control groups. Subreddit bans, shown in the left column, illustrate the overall activity of users comparing the year before the subreddit ban to the year after. Recall that we define active users as those that posted at least 10 times in the pre-ban period. Because no new users can join a banned subreddit we expect to see a natural decrease in the average weekly activity for the treatment group as users leave the platform.

\begin{table}[t]
    \centering
        \caption{Results of difference in differences (DiD) analysis comparing treatment groups to each control. Bold indicates statistically significant decrease (one-tailed $p<0.05$).}
    \scriptsize{
    \begin{tabular}{r l || r l | r l}
        & & \multicolumn{2}{c}{\scriptsize{Control A}} & \multicolumn{2}{c}{\scriptsize{Control B}} \\
        & \textbf{Treatment Group} & \textbf{DiD} & $p$-value &  \textbf{DiD} & $p$-value  \\ \hline
        \multirow{8}{*}{\rotatebox[origin=c]{90}{Regulatory}} & r/Physical\textunderscore Removal & \textbf{-7.47} & 0.0001 & \textbf{-17.19} &  0.0001 \\
        & r/GunsForSale  & 0.00 & 0.9190 & -0.03 & 0.9230 \\
        & r/SanctionedSuicide  & \textbf{-2.47} & 0.0052  & \textbf{-2.58} & 0.0057 \\
        & r/DarkNetMarkets  & \textbf{-3.59} & 0.0002  & \textbf{-5.58} & 0.0004 \\
        & r/Incels & \textbf{-1.02} & 0.0070 & \textbf{-3.57} & 0.0070\\
        & r/fatpeoplehate  &  \textbf{-3.23} & 0.0005 & \textbf{-4.10} & 0.0005\\
        & r/greatawakening & \textbf{-8.14} & 0.0001  & \textbf{-9.09} & 0.0001  \\
        & r/MillionDollarExtreme  & \textbf{-8.40} & 0.0001 & \textbf{-9.75} & 0.0001 \\ \hline
        \multirow{8}{*}{\rotatebox[origin=c]{90}{External}} & r/The\textunderscore Donald (RNC)  &  \textbf{-0.69}  &  0.0001 & \textbf{-0.92} & 0.0009 \\
        & r/The\textunderscore Donald (Election) & 0.68 &  0.6930 & 0.56 & 0.6930 \\
        & r/SandersForPresident  & \textbf{-2.29} & 0.0001 & \textbf{-1.71} & 0.0002 \\
        & r/StLouisRams  & \textbf{-5.47} & 0.0001 & \textbf{-4.52} & 0.0001 \\
        & r/Chargers & 0.60 & 0.1809  & 0.45 & 0.1774 \\
        & r/StarWars (Ep7) & 0.66 &  0.1025  & -0.06 & 0.1036 \\
        & r/StarWars (Ep8)  & 0.03 &  0.1502 & -0.23 & 0.1456 \\
        & r/DarkNetMarkets (BTC) & \textbf{-2.03} & 0.0005 & \textbf{-3.79} & 0.0005 \\
    \end{tabular}
    }
    \label{tab:did}
\end{table}

In many cases, however, the difference in pre-ban growth compared to post-ban activity is stark, which indicates that the ban had a chilling effect on the overall posting behavior of the affected users. An important caveat for the analysis of regulatory actions (left column) is that access to a banned subreddit's posts and comments immediately becomes unavailable to our data collection approach in the event of a ban. Although we continue to collect active user activity from other subreddits, we were sometimes unable to collect data for the banned subreddit. This missing data would cause a misleading illustration, so we remove data points for the weeks where we were not able to collect data. For example, in the r/MillionDollarExtreme plot (bottom-left plot in Fig~\ref{fig:activity}) we were not able to collect data for the seven weeks prior to the ban; hence, seven weeks of treatment-group data is not plotted.

\subsubsection{Difference in Differences Analysis.}

Recall that control groups A and B are used to compare the relative change of the treatment group. Although the results illustrated in Fig.~\ref{fig:activity} are visually compelling, we require a more rigorous statistical test to understand whether these activity changes are statistically significant compared to the controls. In order to account for unmeasured time-invariant confounders we performed a difference in differences (DiD) analysis. The DiD test measures the difference in the effect of a treatment on the treatment group versus the effect on the control groups. Presumably, the treatment effect on the control groups should be zero, so any effect found in the treatment group should be subtracted from the effect found in the controls. Again consider, as an example, the r/MillionDollarExtreme measurements in the bottom left plot of Fig.~\ref{fig:activity}. Here the large drop in post-ban activity of active users is not matched by a post-ban drop in the controls.

We calculate the mean-average activity in the treatment group (except for weeks with missing data) and control groups pre-ban and post-ban. $\Delta_\textrm{pre}$ is the difference between the mean pre-ban treatment and the control activity; $\Delta_\textrm{post}$ is the difference between the mean post-ban treatment and the control activity. Finally, DiD is $\Delta_{post}-\Delta_{pre}$. Therefore, the DiD analysis calculates the effect of the treatment on the post-ban behavior by comparing the average change in the treatment variable compared to the average change in the controls. Table~\ref{tab:did} presents the results of the DiD analysis. Negative DiD results indicate that the ban or external event reduced user activity. Conversely, positive DiD indicate that the ban or external event increased user activity.

Our next analysis is to determine if these effects were due to chance. Unfortunately, there do not exist off-the-shelf significance testing for DiD analysis like a t-test. Instead we perform permutation tests to elicit a $p$-value for each treatment group. To perform these tests, we simulated thousands of treatment-subreddits from among the potential control-users. Specifically, for each treatment-group we collected the posting histories of all the unmatched active users for each of the 200 subreddits in control group A, \ie, subreddits that were similar to the treatment subreddits. Then we randomly sampled and combined these users to create simulated treatment subreddits. DiD tests were completed between each simulated subreddit to create a distribution of DiD results. This distribution was then compared to the actual DiD results. The $p$-value indicates the percentage of the simulated DiD results that were lower than the actual DiD result. Bold DiD results in Table~\ref{tab:did} indicate when the DiD effects are significant at the 0.05 level.

\begin{figure}[t]
    \centering
    \input{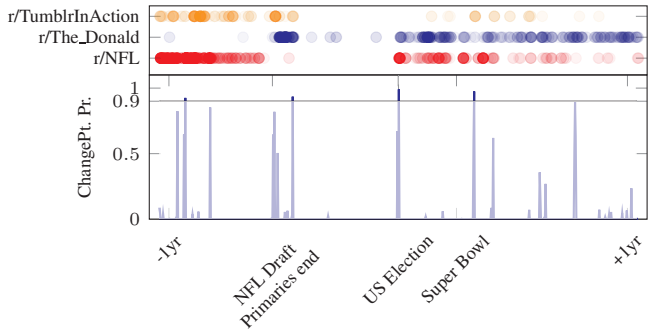}
    \caption{Example changepoint analysis for a randomly selected user from r/The\textunderscore Donald (US Election) treatment group centered on the 2016 US presidential election (Nov 8, 2016). Changepoint analysis assigns the probability that a behavior change occurred at each point in time. Top plot illustrates the number of posts and comments for three of the subreddits frequented by this user. Bottom plot shows the corresponding changepoint probability represented by a blue line; the 90\% threshold is met 4 times by this user.}
    \label{fig:cp_example}
\end{figure}

\begin{figure}[t!]
\input{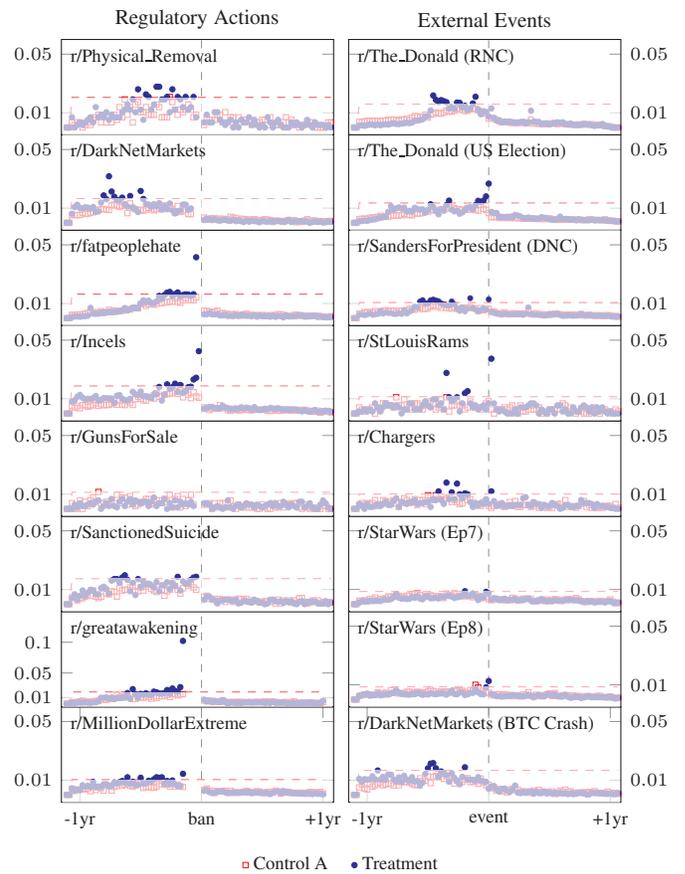}
\caption{Probability that a user in the treatment group (\textcolor{blue}{\Circle}) or control group A (\textcolor{red}{\Square}) changes their behavior in each week. We observe that changepoints with $p(y_i)\ge0.9$ are relatively rare. The red dashed line is placed at 3 standard deviations away from the mean number of changepoints in the control group. The probability that a user experiences a changepoint typically spikes near the subreddit ban or external event date indicating statistically significant change in average behavior at these highlighted points.  
}
\label{fig:changepoint}
\end{figure}

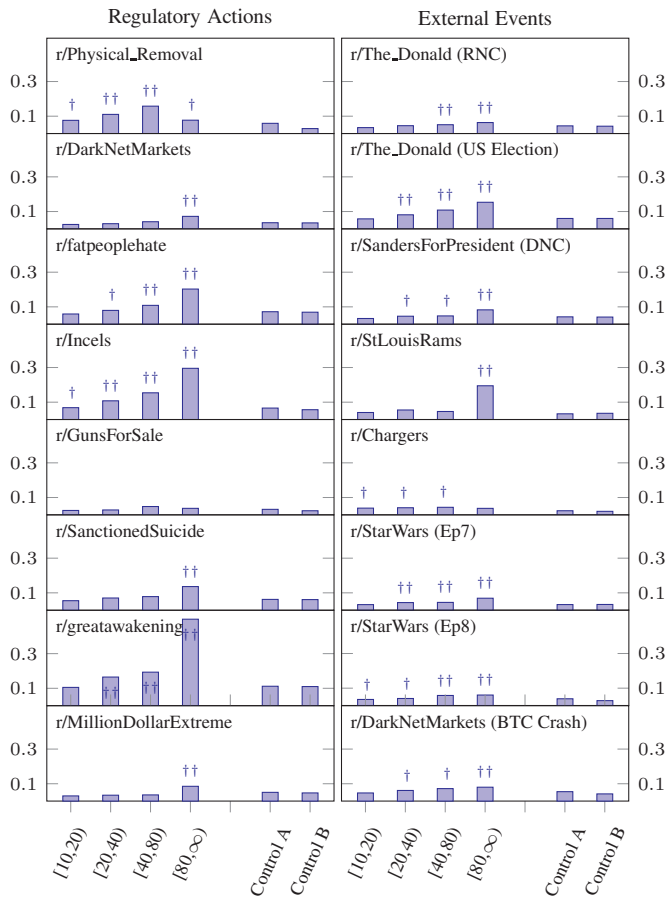
\begin{figure}
    \begin{tikzpicture}
\begin{groupplot}[group style={group size=2 by 8,  horizontal sep=0.1cm, vertical sep=0cm },height=2.85cm,width=5.4cm,ymin=0, tick label style={font=\scriptsize}, every axis title/.style={below right,at={(0,1)},font=\scriptsize}, scaled y ticks=false, ]
\nextgroupplot[
        ybar, 
        bar width=6pt, 
        title=r/Physical\_Removal,
        ymax=0.55, 
        ytick={0.1,0.3},
        xtick=\empty
]

\addplot+[nodes near coords, nodes near coords style={above}, point meta=explicit symbolic] coordinates
	{
(1,0.0760456273764259)[{\tiny~$\dag$}] 
(2,0.11038961038961)[{\tiny$\dag\dag$}]
(3,0.157894736842105)[{\tiny$\dag\dag$}]
(4,0.0769230769230769)[{\tiny~$\dag$}]
(6,0.0588235294117647)
(7,0.0285714285714286)
};

\coordinate (c1) at (rel axis cs:0,1);

\nextgroupplot[
        ybar, 
        bar width=6pt, 
        title=r/The\_Donald (RNC),
        ymax=0.55, 
        ytick={0.1,0.3},
        xtick=\empty,
        yticklabel pos=right
]

\addplot+[nodes near coords, nodes near coords style={above}, point meta=explicit symbolic] coordinates
{
(1,0.0341374960580259)
(2,0.0452204496922497)
(3,0.050834740198837)[{\tiny$\dag\dag$}]
(4,0.0633225726347157)[{\tiny$\dag\dag$}]
(6,0.0442137977541267)
(7,0.0422617653945212)
};

\coordinate (c2) at (rel axis cs:1,1);

\nextgroupplot[       
        ybar,
        bar width=6pt, 
        title=r/DarkNetMarkets,
        ymax = 0.55, 
        ytick={0.1,0.3},
        xtick=\empty
]


\addplot+[nodes near coords, nodes near coords style={above}, point meta=explicit symbolic] coordinates
	{
	(1,0.0253784505788068)
(2,0.029796511627907)
(3,0.0411311053984576)
(4,0.0720338983050848)[{\tiny$\dag\dag$}]
(6,0.0353930387172468)
(7,0.0343817151787459)
};

\nextgroupplot[       
        ybar,
        bar width=6pt, 
        title=r/The\_Donald (US Election),
        ymax=0.55, 
        ytick={0.1,0.3},
        xtick=\empty,
        yticklabel pos=right
]

\addplot+[nodes near coords, nodes near coords style={above}, point meta=explicit symbolic] coordinates
{
(1,0.0576073865917302)
(2,0.0811029411764706)[{\tiny$\dag\dag$}]
(3,0.108275648310776)[{\tiny$\dag\dag$}]
(4,0.153347732181425)[{\tiny$\dag\dag$}]
(6,0.0599564077240064)
(7,0.0600166902259993)
};

\nextgroupplot[
        ybar, 
        bar width=6pt, 
        title=r/fatpeoplehate,
        ymax=0.55, 
        ytick={0.1,0.3},
        xtick=\empty
]


\addplot+[nodes near coords, nodes near coords style={above}, point meta=explicit symbolic] coordinates
	{
(1,0.0587927424982554)
(2,0.0795309711955136)[{\tiny$~\dag$}]
(3,0.108573717948718)[{\tiny$\dag\dag$}]
(4,0.202655485674354)[{\tiny$\dag\dag$}]
(6,0.0719154086247532)
(7,0.0691232528589581)
};

\nextgroupplot[
        ybar, 
        bar width=6pt, 
        title=r/SandersForPresident (DNC),
        ymax=0.55, 
        ytick={0.1,0.3},
        xtick=\empty,
        yticklabel pos=right
]

\addplot+[nodes near coords, nodes near coords style={above}, point meta=explicit symbolic] coordinates
{
(1,0.0328228944395632)
(2,0.0459675764416171)[{\tiny$~\dag$}]
(3,0.0476897689768977)[{\tiny$~\dag$}]
(4,0.0826666666666667)[{\tiny$\dag\dag$}]
(6,0.0423396902598234)
(7,0.0410902861233987)
};

\nextgroupplot[
        ybar,
        bar width=6pt, 
        title=r/Incels,
        ymax=0.55, 
        ytick={0.1,0.3},
        xtick=\empty
]

\addplot+[nodes near coords, nodes near coords style={above}, point meta=explicit symbolic] coordinates
{
(1,0.0683353198251887)[{\tiny$~\dag$}]
(2,0.107843137254902)[{\tiny$\dag\dag$}]
(3,0.154037267080745)[{\tiny$\dag\dag$}]
(4,0.295373665480427)[{\tiny$\dag\dag$}]
(6,0.0660181789188327)
(7,0.0565828986881618)
};

\nextgroupplot[
        ybar, 
        bar width=6pt, 
        title=r/StLouisRams,
        ymax=0.55, 
        ytick={0.1,0.3},
        xtick=\empty,
        yticklabel pos=right
]

\addplot+[nodes near coords, nodes near coords style={above}, point meta=explicit symbolic] coordinates
{
(1,0.0404411764705882)
(2,0.0549450549450549)
(3,0.0462962962962963)
(4,0.194690265486726)[{\tiny$\dag\dag$}]
(6,0.0328571428571429)
(7,0.0358166189111748)
};

\nextgroupplot[
        ybar, 
        bar width=6pt, 
        title=r/GunsForSale,
        ymax=0.55, 
        ytick={0.1,0.3},
        xtick=\empty
]

\addplot+[nodes near coords, nodes near coords style={above}, point meta=explicit symbolic] coordinates
{
(1,0.0253411306042885)
(2,0.028169014084507)
(3,0.048)
(4,0.0375)
(6,0.0317460317460317)
(7,0.0236220472440945)
};

\nextgroupplot[
        ybar, 
        bar width=6pt, 
        title=r/Chargers,
        ymax=0.55, 
        ytick={0.1,0.3},
        xtick=\empty,
        yticklabel pos=right
]

\addplot+[nodes near coords, nodes near coords style={above}, point meta=explicit symbolic] coordinates
{
(1,0.0387453874538745)[{\tiny$\dag~$}]
(2,0.040983606557377)[{\tiny$\dag~$}]
(3,0.0434782608695652)[{\tiny$\dag~$}]
(4,0.0374149659863946)
(6,0.023680649526387)
(7,0.0202292650033715)
};

\nextgroupplot[
        ybar, 
        bar width=6pt, 
        title=r/SanctionedSuicide,
        ymax=0.55, 
        ytick={0.1,0.3},
        xtick=\empty
]

\addplot+[nodes near coords, nodes near coords style={above}, point meta=explicit symbolic] coordinates
{
(1,0.0547073791348601)
(2,0.0702127659574468)
(3,0.0782918149466192)
(4,0.136186770428016)[{\tiny$\dag\dag$}]
(6,0.062190423775454)
(7,0.0610225398570643)
};

\nextgroupplot[
        ybar, 
        bar width=6pt, 
        title=r/StarWars (Ep7),
        ymax=0.55, 
        ytick={0.1,0.3},
        xtick=\empty,
        yticklabel pos=right
]

\addplot+[nodes near coords, nodes near coords style={above}, point meta=explicit symbolic] coordinates
{
(1,0.0320690216506593)
(2,0.0438043804380438)[{\tiny$\dag\dag$}]
(3,0.0448678549477566)[{\tiny$\dag\dag$}]
(4,0.0689922480620155)[{\tiny$\dag\dag$}]
(6,0.0323329331732693)
(7,0.032914230800032)
};

\nextgroupplot[
        ybar, 
        bar width=6pt, 
        title=r/greatawakening,
        ymax=0.55, 
        ytick={0.1,0.3},
        xtick=\empty
]

\addplot+[nodes near coords, nodes near coords style={below}, point meta=explicit symbolic] coordinates
{
(1,0.1052378664)
(2,0.1648793566)[{\tiny$\dag\dag$}]
(3,0.1931097008)[{\tiny$\dag\dag$}]
(4,0.4986684421)[{\tiny$\dag\dag$}]
(6,0.1118421053)
(7,0.110056926)
};

\nextgroupplot[
        ybar, 
        bar width=6pt, 
        title=r/StarWars (Ep8),
        ymax=0.55, 
        ytick={0.1,0.3},
        xtick=\empty,
        yticklabel pos=right
]

\addplot+[nodes near coords, nodes near coords style={above}, point meta=explicit symbolic] coordinates
{
(1,0.0348921234644369)[{\tiny$~\dag$}]
(2,0.0404794655138534)[{\tiny$~\dag$}]
(3,0.0579055441478439)[{\tiny$\dag\dag$}]
(4,0.0603580562659847)[{\tiny$\dag\dag$}]
(6,0.0388188281695988)
(7,0.0280945043656908)
};

\nextgroupplot[
        ybar, 
        bar width=6pt, 
        title=r/MillionDollarExtreme,
        xtick = {1,2,3,4,5,6,7},
        xticklabels = {{[10,20)},{[20,40)},{[40,80)},{[80,$\infty$)}, {~} ,Control A,Control B},
        ymax=0.55,
        ytick={0.1,0.3},
        xticklabel style={rotate=65},
        legend style={at={($(0,0)+(1cm,1cm)$)}, nodes={scale=0.7, transform shape} , legend columns=4,fill=none,draw=none,anchor=center,align=center},
            legend to name=fred
]

\addplot+[nodes near coords, nodes near coords style={above}, point meta=explicit symbolic] coordinates
{
(1,0.02976567448)
(2,0.03392568659)
(3,0.03553800592)
(4,0.08522412839)[{\tiny$\dag\dag$}]
(6,0.05035460993)
(7,0.04701307883)
};

\nextgroupplot[
        ybar, 
        bar width=6pt, 
        title=r/DarkNetMarkets (BTC Crash),
        xtick = {1,2,3,4,5,6,7},
        xticklabels = {{[10,20)},{[20,40)},{[40,80)},{[80,$\infty$)}, {~} , Control A,Control B},
        ymax=0.55,
        ytick={0.1,0.3},
        xticklabel style={rotate=65},
        yticklabel pos=right
]

\addplot+[nodes near coords, nodes near coords style={above}, point meta=explicit symbolic] coordinates
{
(1,0.0468364831552999)
(2,0.0612648221343874)[{\tiny~$\dag$}]
(3,0.0719424460431655)[{\tiny~$\dag$}]
(4,0.0800477897252091)[{\tiny$\dag\dag$}]
(6,0.0542786421499293)
(7,0.0413901501921062)
};

\end{groupplot}
    \coordinate (c3) at ($(c1)!.5!(c2)$);
    \node[below] at (c3 |- current bounding box.south)
      {\pgfplotslegendfromname{fred}};
      
\node (title) at ($(group c1r1.center)+(0,0.9cm)$) {\footnotesize{Regulatory Actions}};
\node (title2) at ($(group c2r1.center)+(0,0.9cm)$) {\footnotesize{External Events}};
\end{tikzpicture}
    \caption{\hl{Probability that a user of a subreddit, posting $x$-times, experienced a changepoint in the month before or after the shock. Left and right $\dag$s over each bar indicate treatment subreddits with changepoint rates that are significantly higher (one-tailed $p<0.05$) than the attrition rates of Control A and Control B respectively. For r/greatawakening and r/MillionDollarExtreme, 2 months of data was collected prior to the shock to discover changepoints because of data collection limitations.}}
    \label{fig:changepoint_probability}
\end{figure}

\subsubsection{Changepoint Analysis}

Next, we employ changepoint detection to look for changes that occur in the trend and amount of activity of the treatment groups. Specifically, we use multivariate Bayesian changepoint analysis to identify when users change their online posting behavior~\cite{xuan2007modeling,adams2007bayesian}. 
Changepoint analysis takes a sequence of observations $y_1, y_2, \ldots, y_T$ corresponding to measurements in a time series then divides these observations into $K$ partitions $\pi_1,\ldots, \pi_K$ such that the data across partitions is independent: $p(y_{1:T} | \pi) = \prod_{k=1}^{K}{p(y_{\pi_k})}$.

Applying these mechanics to our analysis, each observation $y_i\in \mathbb{R}^d$ is a $d$-dimensional vector representing the number of posts and comments made by a user in each of the $d$ subreddits in the $i^\textrm{th}$ week before or after the ban or external event date. 
We use the full covariance model from Adams and MacKay\footnote{We reimplemented the original code and make it available at \url{http://github.com/danjr26/changepoint-analysis}.} to represent $y_i$ as a multivariate Poisson, $y_i \sim \textrm{Pois}(\lambda_{0}, \lambda_{1}, \ldots, \lambda_{d})$~\cite{adams2007bayesian}, where $\lambda_{x}$ is the mean activity rate for the user in a subreddit. Simply put, the changepoint analysis outputs the probability that a user's posting behavior (\ie, their posting frequency and/or which subreddit they posted to) changed in the $i^\textrm{th}$ week. We use a threshold of $p(y_i)\ge 0.90$ to draw partition boundaries.

The example in Fig.~\ref{fig:cp_example} plots the changepoint of an example user who posted to r/nfl, r/The\textunderscore Donald, and r/TumblrInAction more than 20 times each (as well as dozens of other subreddits that are not illustrated to maintain clarity). We observe that the changepoint probability spikes above the 90\% threshold four times: (1) a year before the election corresponding to changes not illustrated in the top plot, (2) \hl{decreased posting activity in r/NFL, approximately beginning with the end of the NFL draft on April 30, 2016, and increased activity in r/The\_Donald, which corresponds to when he effectively clenched the nomination after the 2016 Indiana Republican primary on May 3, 2016}, (3) immediately after the US Election which corresponds to a large increase in activity, and (4) 116 days after the US Election corresponding to a large increase in posts to r/TumblrInAction, \hl{and a decrease in activity in r/NFL after the 2017 Super Bowl}. \hl{Of course, this user is not representative of all Reddit users, but is rather meant to showcase an example of how a changepoint analysis works.  In this case, the example user experienced 4 changepoints at the times given above.} The 90\% probability used in this work was chosen to be a high threshold generalizable across users; It is not our intention to perform an exhaustive analysis of changepoint thresholds although further research to explore the changepoint threshold may yield interesting results. \hl{How users are changing, \ie, which subreddits they start or stop posting in, is out of the scope of this work.  Simply put, we are currently only interested in \textit{if} a change occurred, not the nature of the change.}

If subreddit bans and important external events significantly affect user behavior, then we should see an increase in the number of users with changepoints identified at or near the ban or event date relative to the controls. Figure~\ref{fig:changepoint} illustrates the probability that users in the treatment group and control group A experience a changepoint in the year before and after the ban or event; control group B is not shown in this illustration for clarity, but is very similar to control group A. Note that, like in Fig.~\ref{fig:activity}, Fig.~\ref{fig:changepoint} has gaps in some of the banned subreddits because that data was deleted before it could be collected. So, although the significant changepoints in r/greatawakening and r/MillionDollarExtreme are not exactly on the ban date, they are as close as the data allows, and appear to occur approximately when the data for the banned subreddits stopped being collected. 

We find that changepoints are relatively rare, with usually less than 2 percent of users undergoing a changepoint in any given week. The upper-bound of \hl{3 standard deviations away from the mean} in control group A is drawn as a horizontal dashed line in Fig.~\ref{fig:changepoint}; we observe that the weeks surrounding the ban or external event do experience a statistically significant uptick in changepoints. \hl{While in our previous analysis we saw mixed results on the effect of external events, here we do find that changepoints exists on or near the dates of most shocks.  This might correspond to an uptick in activity in these subreddits as users gather to discuss their NFL team moving or the release of the new Star Wars movie. However, we can conclude that there are users who do change their behavior as the result of both external and regulatory shocks.}

In some cases, changepoints did not align with our initial expectations. For example, the r/StLouisRams and r/Chargers also had changepoints that correspond with the start of the NFL season, as well as the end of the season when the moves were announced. \hl{These changes still correspond to significant external events, just not the events we initially hypothesized.  Additionally, we see that many communities have many changepoints outside of the shock.  This suggests that users who are posting in these communities might be more volatile in their Reddit posting behavior compared to the control. This may be because these users are joining and leaving new communities more frequently, or varying their posting behavior.  More research is needed to uncover why these community members are experiencing changepoints.}

\hl{We also find that the level of engagement in the community increases the likelihood that a user will change their behavior after a shock. Figure~\ref{fig:changepoint_probability} illustrates the probability that a user will experience a changepoint in the month before or after the shock, given that they posted between between 10 and 20, 20 and 40, 40 and 80, or over 80 times respectively. In almost all cases, indicated by left and right $\dag$, users who are more active in a subreddit are significantly more likely to experience a change compared to Control A and Control B respectively (one-tailed $p$-value $< 0.05$)}

Finally, it is important to note that these changepoints capture all posting and commenting activity on any subreddit. Our changepoint analysis does not indicate precisely what changed -- only that something changed. Further research is needed to understand this in more detail.

\section{Predicting Attrition After a Shock}

The previous analysis showed that subreddit bans and certain external events have a small but statistically significant effect on user behavior. Our next goal is to \hl{determine if there is enough information to predict if a user will stay or leave Reddit after the event (RQ3)}.

\begin{table}[t]
    \centering
        \caption{User attrition modelling results for user attrition (RQ3) under cross validation and transfer scenarios. }
    \scriptsize{
    \begin{tabular}{r l || c c | c c }
        & & \multicolumn{4}{c}{Attrition}  \\
        & & \multicolumn{2}{c|}{Cross Validation} & \multicolumn{2}{c}{Transfer}  \\
        & \textbf{Subreddit} & \textbf{AUC} & \textbf{F1} & \textbf{AUC}  & \textbf{F1}  \\ \hline
        \multirow{8}{*}{\rotatebox[origin=c]{90}{Regulatory}} & r/Physical\textunderscore Removal & {0.62$\pm$0.027} & 0.79 $\pm$ 0.024 & 0.63 & 0.81 \\
        & r/GunsForSale  & {0.78$\pm$0.021}& 0.90 $\pm$ 0.010 & 0.78 & 0.90  \\
        & r/SanctionedSuicide  & {0.73$\pm$0.025}& 0.68 $\pm$ 0.019& 0.71 & 0.66  \\
        & r/DarkNetMarkets & {0.79$\pm$0.011}& 0.67 $\pm$ 0.010& 0.74 & 0.64  \\
        & r/Incels & {0.76$\pm$0.011}& 0.76 $\pm$ 0.014 & 0.73 & 0.74  \\
        & r/fatpeoplehate  &  {0.80$\pm$0.006}& 0.80 $\pm$ 0.006 & 0.79 & 0.80  \\
        & r/greatawakening  & 0.81$\pm$0.015& 0.72 $\pm$ 0.019& 0.77& 0.61  \\
        & r/MillionDollarExtreme  & 0.69$\pm$0.017 & 0.73 $\pm$ 0.006 & 0.69 & 0.72 \\ \hline
        \multirow{8}{*}{\rotatebox[origin=c]{90}{External}} & r/The\textunderscore Donald (RNC) &  0.70$\pm$0.006 & 0.91 $\pm$ 0.002 & 0.67 & 0.91 \\
        & r/The\textunderscore Donald (Election) &  0.71$\pm$0.005& 0.90 $\pm$ 0.002 & 0.67 & 0.91  \\
        & r/SandersForPresident &  0.79$\pm$0.006 & 0.92 $\pm$ 0.001 & 0.71 & 0.92 \\
        & r/StLouisRams &  0.63$\pm$0.067 & 0.95 $\pm$ 0.017 & 0.62 & 0.95  \\
        & r/Chargers &  0.72$\pm$0.089 & 0.96 $\pm$ 0.005& 0.66 & 0.96  \\
        & r/StarWars (Ep7) &  0.69$\pm$0.023 & 0.97 $\pm$ 0.002 & 0.69 & 0.97 \\
        & r/StarWars (Ep8)  &  0.69$\pm$0.014& 0.96 $\pm$ 0.001 & 0.66 & 0.96 \\
        & r/DarkNetMarkets (BTC) &  0.73$\pm$0.014 & 0.73 $\pm$ 0.017& 0.62 & 0.72  \\
    \end{tabular}
    }
    \label{tab:leavestay}
\end{table}

\subsubsection{Attrition Results}
 First, we use standard 5-fold cross validation on the treatment group. In this case, the active users of the treatment subreddit are randomly split into 5 equal-sized sets. In each fold, four sets are combined to create a training set and one set is held out for validation. 

We experimented with several different predictive models. Ultimately we chose a straightforward two-layer neural network model, but resisted the temptation to highly optimize hyper-parameters and the model architecture.

From the treatment group users, we identified the $m$ subreddits to which they contributed at least one post or comment. Then, for each user, the input neurons $x_1,x_2,\ldots,x_m$ were set to contain the total number of posts and comments made in each subreddit in the 6 months prior to the ban or external event. The input layer is fully connected to two hidden layers with 1000 neurons each with a RELU activation function. The second hidden layer is fully connected to two output nodes $y, \bar{y}$ containing a sigmoid activation function. The output layer indicates the likelihood that the user stays on Reddit or does not respectively after a 4-week grace period. We used log mean squared error (LMSE) for the loss function and assigned a prediction label (stay or leave) based upon the output node with the highest weight. 

Training is performed for each treatment group independently under both cross validation and transfer scenarios. Results and 95\% confidence intervals for the cross validation results are presented in Table~\ref{tab:leavestay}. We find that our model can predict user attrition with modest AUC in both the 5-fold cross validation and transfer scenarios. This suggests that it is possible to predict which users will stay leave the platform, given past posting history, with fair accuracy.

\section{Conclusions}

\hl{We conclude by revisiting the research questions asked at the beginning of this paper.}

\noindent\hl{\textbf{RQ1: How do online communities react to different kinds of shocks?}
Although previous work has shown that subreddit bans are very effective at removing hateful content from Reddit~\cite{chandrasekharan2017you}, we find that subreddit bans and external events typically result in only a small decrease in the number of users and number of posts per user. Combined, these results suggest that subreddit bans may be able to refocus users' attention towards producing non-hateful content rather than simply removing them from the site. For those that do make a change, it remains unclear how they change.  
}

\hl{
We also observe several interesting changes in activity related to external events.
For example, members of the r/SandersForPresident subreddit have a larger decrease in activity compared to the r/The\textunderscore Donald, due to Sanders' failed candidacy and Trump's nomination.
In the r/StLouisRams subreddit have far less activity after the move to Los Angeles than the r/Chargers subreddit. It is likely that, by keeping the r/Chargers subreddit name and identity, there existed a coherent space for users to continue discussing football and their teams' move. 
Additionally, we see that before the teams announced their move, active members of r/Chargers and r/StLouisRams community both had changepoints that correspond to the start of the NFL season.}


\hl{
The Bitcoin crash also had a large external effect on r/DarkNetMarkets, as illustrated in Figure~\ref{fig:leaving}. This result suggests that the Bitcoin crash did have a significant impact on the amount of time that those users spent discussing Bitcoin. After r/DarkNetMarkets was banned, just a few months later, active users decreased their overall posting behavior even further.}

\noindent\hl{\textbf{RQ2: How do individual users react to these shocks?}
Despite these statistically significant findings, we were surprised to observe that most users continued their usual posting behavior. Changepoint analysis found that most users did not change their behavior. The largest effect occurred during the ban of r/greatawakening, in which  10.2\% of affected users had a changepoint registered. In most communities, less than 5 percent of users experienced a changepoint as a result of the shock.}

%

\noindent\hl{\textbf{RQ3: How many users change their behavior as a result of these shocks?}
We observed that a small, but statistically significant, number of active users will leave Reddit after their subreddit is banned. Unsurprisingly, these effects are higher if the user contributed more frequently to the banned subreddit. However, the large majority (typically 90\% or more) of active users who were active in these banned subreddits do stay on the platform -- even after accounting for a generous grace period. These results suggest that even though Reddit bans do rid the platform of many of the dedicated members of these offensive subreddits, as seen in Figure~\ref{fig:leaving}, most stay around despite the ban. We also observe, in most cases, a small, but statistically significant, reduction in the number of posts and comments made by those active users of a banned subreddit.}

\hl{It is important to note that the present work is limited by its selection of subreddits and events. Although there are only a handful of popular subreddits that have been banned, there are indeed many subreddits that have experienced external shocks. We only sample a few in the present work. However, the methods used in this paper constitute an expanded framework that can be used to study online communities in future work.} 

\hl{It is also important to note that our analysis only captures the \textit{amount} and \textit{frequency} of posts and comments that are made on Reddit. We do not consider the words used in each activity, nor do we observe the subreddits that people are browsing, or the posts and comments they upvote or downvote. As is typical in Web culture, most social media users are passive viewers of content~\cite{glenski2017consumers}, so we do not claim to capture all user activity. However, by using post and comment activity history, the present work can understand the effect that external events and subreddit bans have on user activity.}

  \section*{Acknowledgments}
This work is sponsored by a grant the US Army Research Office (W911NF-17-1-0448) and a grant from USAID DRG Advancing Media Integrity Program (7200AA18CA00059).



\bibliographystyle{abbrv}
\bibliography{IEEEabrv,bibliography.bib}

\end{document}